*Comment on **Global dynamics of biological systems from time-resolved omics experiment.** Bioinformatics. 2006 Jun 15; 22(12):1424-30.*
*Radhakrishnan Nagarajan; rnagarajan@uams.edu*
*Department of Biostatistics, University of Arkansas for Medical Sciences*

In a recent study, (Grigorov, 2006) analyzed temporal gene expression profiles (Arbeitman et al., 2002) generated in a Drosophila experiment using SSA in conjunction with Monte-Carlo SSA. The author (Grigorov, 2006) makes three important claims in his article, namely:

> **Claim 1**: A new method based on the theory of *nonlinear time series* analysis is used to capture the global dynamics of the fruit-fly cycle temporal gene expression profiles.
>
> **Claim 2**: Flattening of a significant part of the eigen-spectrum confirms the hypothesis about an *underlying high-dimensional chaotic generating process*.
>
> **Claim 3**: Monte-Carlo SSA can be used to establish whether a given time series is distinguishable from any well-defined process including *deterministic chaos*.

In this report we present fundamental concerns with respect to the above claims (Grigorov, 2006) in a systematic manner with simple examples. The discussion provided especially discourages the choice of SSA for inferring nonlinear dynamical structure form time series obtained in any biological paradigm.

1. The technique used by the author (Grigorov, 2006) relies on singular-spectrum analysis (SSA). While the author has claimed to have used a *nonlinear time series* approach, SSA by very definition is a *linear decomposition technique.* The lag-correlation matrix constructed as a part of the SSA technique is a measure of *linear correlation between the samples* in the time series, hence *representative of only linear statistical properties in the given time series*. On a related note, SSA is used for non-parametric power-spectral estimation (Caratheodory and Fejer, 1911; Pisarenko, 1973; Sidiropoulos, 2001). *Thus any claims nonlinear dynamical properties in the fruit-fly temporal gene expression profiles based on the results of SSA should be strongly discouraged*. It is equally important to note that the concatenation procedure adopted in (Grigorov, 2006) does not necessarily constitute a time series analysis. A time series is basically discrete representation of a continuous signals sampled in amplitude and time. The sampling in turn can be uniform or non-uniform. The time axis fails to make sense if such time series are concatenated. In the present paradigm mere visualization of the gene expression profiles reveals that they share no apparent statistical similarities. Thus the concatenation procedure even if termed as a time series is likely to be non-stationary.

2. Flattening of the eigen-spectrum of the lag-correlation matrix determined using SSA *does not confirm the presence of high-dimensional chaotic generating process*. As noted earlier, SSA is a linear decomposition technique. The lag-correlation matrix constructed as a part of the SSA procedure is a measure of only *linear correlations* in the time series. While SSA is useful as a spectral estimation tool *it cannot provide any insight into the dimensionality of nonlinear dynamical system*. It might not be surprising to note that nonlinear dynamical systems with widely different dimensionalities can have the same eigen-spectrum estimated by SSA. The possible confusion in (Grigorov, 2006) may have been caused due to the similarity of the lag-matrix used in spectral estimation (Caratheodory's Uniqueness theorem) (Caratheodory and Fejer, 1911) to the trajectory matrix used in embedding with optimal delay $\tau = 1$ (Takens' Theorem) (Takens, 1981; Mane, 1981; Sauer et al, 1991). It is important to appreciate the fact that the elements of the lag-matrix are linearly correlated whereas those of the trajectory matrix are usually nonlinearly correlated. In this report, an example of a low-dimensional chaotic nonlinear dynamical system from Grigorov, 2006 (Lorenz attractor) is used to elucidate this point. SSA may be useful in capturing the spectral properties of the nonlinear dynamical process but provides no insight into its dimensionality. Flattening of the eigen-spectrum can be observed across two distinct processes with widely varying dynamical properties. A classic example is that of a chaotic logistic map and its phase-randomized surrogate (Theiler et al., 1992; Kantz and Schreiber, 1997). *Thus any conclusion on underlying high-dimensional chaos in the fruit-fly temporal gene expression profiles based on the results of SSA can be challenged*. Techniques such as SSA implicitly assume the given time series to be wide-sense stationary (WSS). Examining the fruit-fly cycle gene expression profiles of the four genes presented in the original article (Arbeitman et al., 2002) reveals no apparent common pattern defying the stationarity constraint. *This in turn discourages "global" analysis of the fruit-fly data by concatenation of the temporal expression profiles as proposed by Grigorov, 2006.*

3. Monte-Carlo SSA as with SSA rely only spectral properties which are insufficient statistics for deterministic chaos. Chaos is an instance of *dynamical nonlinearity* and its properties are captured by *dynamical invariants* (Kantz and Schreiber, 1997). For instance it is not uncommon to encounter a chaotic time series which exhibits a broad-band power-spectrum characteristic of white noise. *Therefore, SSA and Monte-Carlo SSA may not be sufficient to distinguish the given data from a chaotic process.*

**SINGULAR SPECTRUM ANALYSIS (SSA)**
SSA begins by constructing the lag-matrix $x^L$ from the given stationary one-dimensional time series using a chosen lag or window-size ($L$). Subsequently, the dominant components in the given data can be identified by eigen-decomposition of the correlation matrix. The eigen-decomposition can also be accomplished by direct singular value decomposition (SVD) of the lag-matrix $x^L$. An example of the lag-matrix construction is shown below.

**Example 1** The lag-matrix $x^L$ from the one-dimensional time series $x = (x_1,...,x_6)$ for a chosen window-size ($L = 3$) is given by

$$x^L = \begin{bmatrix} x_1 & x_2 & x_3 \\ x_2 & x_3 & x_4 \\ x_3 & x_4 & x_5 \\ x_4 & x_5 & x_6 \end{bmatrix}$$

**SSA AND LINEAR CORRELATION:** The correlation matrix (C = $(x^L)^t \cdot x^L$, is a *measure of linear correlation* between the columns of $x^L$. Interestingly, the dominant eigen-values of the correlation matrix reflect the dominant frequency components in the given data. The theory is well established by Caratheodory's uniqueness theorem (Caratheodory and Fejer, 1911). Related technique used widely in signal processing literature for non-parametric power-spectral estimation is Pisarenko Harmonic Decomposition or subspace decomposition (Pisarenko, 1973). Ideally, the lag ($L$) should be chosen so as to be greater than the periodicity in the given signal.

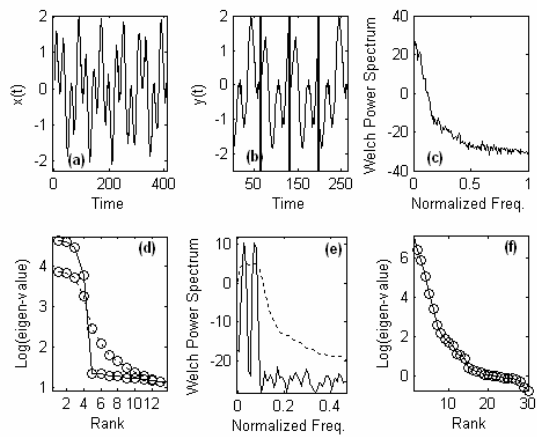

**Fig. 1.** Periodic sine-wave $x(t)$ (Example 2) is shown in Fig. 1a. Sine-wave $y(t)$ obtained by concatenation of four randomly chosen patches of length 66 from $x(t)$ is shown in Fig.1b. The vertical lines in Fig. 1b, separates the four segments. Eigen-spectrum and power-spectrum of $x(t)$ and $y(t)$ are shown by solid and dotted lines in Fig.1d and Fig. 1e respectively. Power-spectrum and eigen-decomposition of the chaotic Lorenz attractor is shown in Figs. 1c and 1d respectively.

**Example 2** Consider a periodic sine-wave with minimal noise given by $x(t) = s(t) + e(t)$ where $s(t) = \sin(2\pi f_1 \omega) + \sin(2\pi f_2 \omega)$ with ($f_1 = 4$, $f_2 = 11$, $\omega = 2\pi F_s$). The sampling frequency ($F_s = 300$) was chosen so as to satisfy the Nyquist criterion, i.e. $F_s > 2.max(f_1, f_2)$. The uncorrelated noise $e(t)$ was generated from a normal distribution with zero-mean and variance 0.01.

Eigen-composition by SSA of $x^L$ corresponding to $x(t)$, Fig. 1a, constructed with window-size ($L = 30$) is shown in Fig. 1d (solid line). The eigen-spectrum in Fig. 1d, exhibits a sharp transition to lower-values after the fourth dominant eigen-value. This is to be expected as $x(t)$ is the superposition of two sinusoidal components reflected by two dominant peaks in the corresponding power-spectrum (Fig. 1e. solid line), each sinusoid in turn being represented by two orthonormal bases functions.

**Remark 1** *Eigen-spectrum in the SSA procedure is estimated from the correlation matrix whose entities represent linear correlations between the rows in $x^L$. Thus only linear statistical properties of the time series can be investigated using SSA. Thus the analysis presented in Grigorov, 2006 cannot provide any insight into the nonlinear dynamical aspects of the fruit-fly temporal gene expression profiles.*

**SSA AND DIMENSIONALITY:** An autonomous nonlinear dynamical system in $\mathbf{R}^M$ can be represented as $d\mathbf{x}/dt = f(\beta, \mathbf{x})$. The function $f$ is a nonlinear function and the qualitative behavior is governed by the process parameter $\beta$. The one-dimensional time series $x_n$, n = 1…N is obtained by sampling a single dynamical variable. The method of delay time embedding )Takens, 1981, Mane, 1981, Sauer et al., 1991) has been shown to be useful in reconstructing an equivalent state space (phase space) in $\mathbf{R}^m$ topologically equivalent to the original state space $\mathbf{R}^M$. This process is called *embedding* and results in a trajectory matrix ($x^T$) whose elements represent points in the phase-space. The choice of optimal embedding dimension ($m$) and optimal time delay ($\tau$) is crucial for proper unfolding of the dynamics in the phase-space. If $d_e$ is the true dimension of the attractor in $\mathbf{R}^m$, then a choice of $m > 2d_e$ has been found to be a sufficient choice for reconstructing the dynamics (Sauer et al., 1991). Several algorithms (Kantz and Schreiber, 1997) have been proposed in the literature for the estimating the optimal embedding dimension (Kennel et al, 1992) and time delay (Liebert and Schuster, 1988) from the given one-dimensional time series data.

**Example 3** Consider a one-dimensional time series $x = (x_1, \ldots, x_6)$. The trajectory matrix $x^T$ representing the dynamics in a ($m = 3$) dimensional space with an optimal delay ($\tau = 1$) is given by

$$x^T = \begin{bmatrix} x_1 & x_2 & x_3 \\ x_2 & x_3 & x_4 \\ x_3 & x_4 & x_5 \\ x_4 & x_5 & x_6 \end{bmatrix}$$

A quick glance would reveal that the trajectory matrix in the embedding procedure $x^T$ resembles the lag-matrix $x^L$ in the SSA procedure (Example 1). *However, this is true only when the optimal delay is ($\tau = 1$) which need not necessarily be true across all nonlinear dynamical systems* (Kantz and Schreiber, 1997) It is important to note that the *rows in $x^T$ represent points in the phase space generated from a chaotic nonlinear dynamical system hence nonlinearly correlated. However, those in $x^L$ are linearly correlated. SSA of $x^T$ as in $x^L$ captures the spectral properties of the chaotic process.* However, it *fails to provide any insight into the dimensionality of the nonlinear dynamical system.*

To elucidate this important point we consider the *low-dimensional chaotic* Lorenz system of equations from (Fig. 1, Grigorov, 2006). This system is represented by coupled nonlinear differential equations with three variables given by

$$\frac{dx}{dt} = \sigma(y-x)$$

$$\frac{dy}{dt} = x(\rho - z) - y$$

$$\frac{dz}{dt} = xy - \beta z$$

By very definition the Lorenz system can be embedded in a *three-dimensional* phase-space. For $(\sigma = 10, \beta = 8/3, \rho = 28)$ this system exhibits chaotic behavior (Fig. 1, Grigorov, 2006). The dimension of the chaotic attractor in the phase space is lesser than three as expected. The power-spectrum and the eigen-spectrum of the chaotic Lorenz time series (Fig. 1c and 1d) and exhibit a *gradual but similar decay which flattens at higher frequencies (rank)*. More importantly, the eigen-spectrum fails to exhibit a sharp transition due it its broad band nature. Therefore, SSA is singularly unhelpful in determining the dimensionality (Mees et al., 1987) of the chaotic Lorenz system. Another classic example is that of time series generated from chaotic logistic map and its phase-randomized surrogate (Theiler, 1992; Kantz and Schreiber, 1997). The eigen-spectrum of these two data sets is similar whereas their phase-space geometry is significantly different (Nagarajan, 2005).

**Remark 2**: *Eigen-spectrum obtained by SSA can provide insight into the spectral properties of the given data but can neither capture the dimensionality nor argue in favor of chaos in a given data. Chaos is an instance of dynamical nonlinearity and SSA by very definition captures only linear statistical properties of the given data. In this report, the chaotic Lorenz attractor had a dimension (< 3). However, the eigen-spectrum exhibited a gradual decay with no sharp transitions failing to reflect the dimensionality. In order to estimate the dimensionality, techniques such as false-nearest neighbors are recommended. A possible explanation for the confusion in Grigorov, 2006 may be attributed to the similarity between the lag-matrix and trajectory matrix for optimal delay ($\tau = 1$). Thus conclusions on underlying high-dimensional chaotic process in the fruit-fly temporal gene expression profiles based on the results from SSA in Grigorov, 2006 may not be valid.*

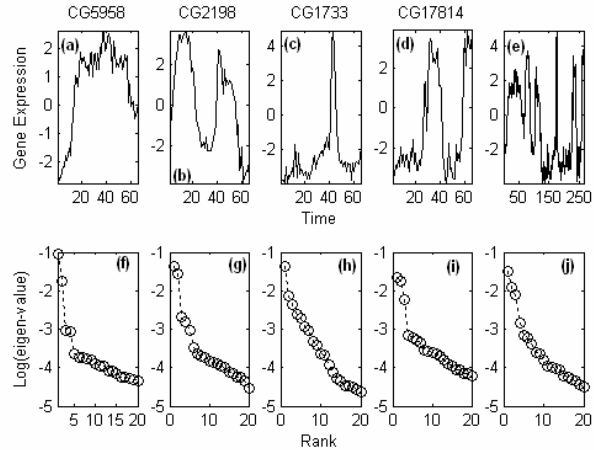

**Fig. 3.** Temporal expression profiles of the four documented genes CG5958, CG2198, CG1733 and CG17184 across 66 time points (Arbeitman et al., 2002) is shown in Figs. 2a-2d respectively. The corresponding concatenated series is shown in Fig. 2e. The eigen-spectrum obtained by SSA is shown right below in Figs. 2f-2j respectively.

**SSA AND STATIONARITY:** SSA implicitly assumes the given signal to be wide-sense stationary. The stationarity constraint can be better appreciated by inspecting the rows of the matrix $x^L$ whose statistical properties are similar for appropriate choice of lag (*L*). As we understand, Grigorov, 2006, appended the expression of the 3104 genes across 66 time points into a single time series and constructed the corresponding lag-matrix $x^L$. As noted earlier, the concatenation procedure destroys the time-ordering hence does not necessarily constitute a time series. To evaluate the performance of such an approach we first considered the stationary time series in Example 2. Four regions each of length 66 were randomly chosen from the periodic sine-wave *x(t)* (Example 2), Fig. 1a, and concatenated. The resulting patchy sine-wave *y(t)* is shown in Fig. 1b. Eigen-decomposition by SSA and power-spectrum of $y(t)$ is shown in Figs. 1d and 1e respectively. The number of dominant eigen-values on SSA of the lag-matrix with window-size ($L = 30$) for $y(t)$ was considerably large than four (Fig. 1d, dotted line). This has to be contrasted with SSA of $x(t)$ whose eigen-spectrum exhibited a sharp transition after the fourth dominant eigen-value. This discrepancy in the eige-spectrum between $x(t)$ and $x(t)$ was also reflected by discrepancies in their power-spectrum, Fig. 1e. Power-spectrum of $y(t)$ failed to exhibit two dominant peaks as expected. There smearing of the spectral power across a range of lower frequencies. Thus concatenating of segments sampled even from stationary processes can result in spurious eigen-values rendering conclusions unreliable.

Inspecting the fruit-fly gene expression profiles for the four documented genes CG5958, CG2198, CG1733 and CG17184) across 66 time points (see Fig. 1D Arbetiman et al., 2006) revealed distinct temporal signatures with no apparent similarity, Figs. 2a-2d. Eigen-spectrum obtained by SSA for each these genes with widow-size ($L = 30$) is shown in Figs. 2f-2i respectively were considerably different across each of these genes indicating no apparent statistical similarity in their correlation/spectral signatures. The time series obtained by concatenation of the temporal profiles of the four genes and the corresponding eigen-spectrum obtained by SSA is shown in Figs. 2e and 2j respectively. The eigen-spectrum of the concatenated series (Fig. 2j) is markedly different from those of the four genes, Figs. 2f-2i. Thus no reliable conclusions can be drawn from SSA analysis of the concatenated time series.

**Remark 3**: *SSA implicitly relies on the wide-sense stationarity (WSS) of the given time series. The rows in the lag-matrix tend to be statistically similar for proper choice of the window-size (L). It might not be prudent to investigate concatenated series using SSA. Such a concatenation can result is significant spectral-smearing even for segments sampled from a stationary periodic process, Fig. 2b. The temporal expression profiles of the four documented genes (Arbeitman et al, 2002, Fig. 1D) were significantly different with distinct eigen-spectra. SSA of the concatenated (Fig. 2j) series shared no resemblance to those of the four genes (Figs. 2f-2i). Therefore, the results of SSA on the fruit-fly gene expression profiles may not lead to meaningful conclusions.*

**MONTE-CARLO SSA** Grigorov, 2006 proposed Monte-Carlo SSA to test whether the given time series was generated from any well-defined process including deterministic chaos. Chaos is an instance of dynamical nonlinearity and chaotic processes are best described by their *invariants* (Kantz and Schreiber, 1997). As noted earlier SSA as well as Monte-Carlo SSA captures only linear statistical properties of the given data which are insufficient to describe nonlinear dynamical processes. Thus Monte-Carlo SSA is singularly unhelpful in distinguishing the given time series from deterministic chaos.